\documentclass[manuscript,screen]{acmart}

\AtBeginDocument{%
  \providecommand\BibTeX{{%
    \normalfont B\kern-0.5em{\scshape i\kern-0.25em b}\kern-0.8em\TeX}}}

\setcopyright{acmlicensed}
\copyrightyear{2024}
\acmYear{2024}
\acmDOI{XXXXXXX.XXXXXXX}

\acmConference[CSCW '24]{27th ACM Conference on Computer-Supported Cooperative Work and Social Computing}{Nov 09-13, 2024}{San José, Costa Rica}
\acmISBN{978-1-4503-XXXX-X/18/06}
\usepackage{soul}
\usepackage{xcolor}
\usepackage{tabularx}
\newcolumntype{b}{>{\raggedright\arraybackslash}X}
\newcolumntype{s}{>{\hsize=.5\hsize}X}
\newcolumntype{u}{>{\hsize=.25\hsize}X}

\begin{document}
\nocite{anonymous}
\title{“We’re not all construction workers”: Algorithmic Compression of Latinidad on TikTok}

\author{Nina Lutz}
\email{ninalutz@uw.edu}
\orcid{0009-0001-8259-8442}
\authornotemark[1]
\email{ninalutz@uw.edu}
\affiliation{%
  \institution{Department of Human Centered Design and Engineering, University of Washington}
  \city{Seattle}
  \state{Washington}
  \country{USA}
}

\author{Cecilia Aragon}
\email{aragon@uw.edu}
\orcid{0000-0002-9502-0965}
\authornotemark[2]
\email{aragon@uw.edu}
\affiliation{%
  \institution{Department of Human Centered Design and Engineering, University of Washington}
  \city{Seattle}
  \state{Washington}
  \country{USA}
}

\renewcommand{\shortauthors}{Lutz and Aragon}

\begin{abstract}
The Latinx diaspora in the United States is a rapidly growing and complex demographic who face intersectional harms and marginalizations in sociotechnical systems and are currently underserved in CSCW research. While the field understands that algorithms and digital content are experienced differently by marginalized populations, more investigation is needed about how Latinx people experience social media and, in particular, visual media. In this paper, we focus on how Latinx people experience the algorithmic system of the video-sharing platform TikTok. Through a bilingual interview and visual elicitation study of 19 Latinx TikTok users and 59 survey participants, we explore how Latinx individuals experience TikTok and its Latinx content. We find Latinx TikTok users actively use platform affordances to create positive and affirming identity content feeds, but these feeds are interrupted by negative content (i.e. violence, stereotypes, linguistic assumptions) due to platform affordances that have unique consequences for Latinx diaspora users. We discuss these implications on Latinx identity and representation, introduce the concept of \textit{algorithmic identity compression}, where sociotechncial systems simplify, flatten, and conflate intersection identities, resulting in compression via the loss of critical cultural data deemed unnecessary by these systems and designers of them. This study explores how Latinx individuals are particularly vulnerable to this in sociotechnical systems, such as, but not limited to, TikTok.

\end{abstract}



\maketitle

\section{Introduction}
Persons of Latin American descent, or Latinx\footnote{While many descriptions of Latine work use phrasing such as “Latino/Hispanic,” this paper purposefully alternates between Latine and Latinx verbiage because of these terms' gender-neutral nature \cite{Scharrón-delRío_Aja_2020, ElCentro_2023} as well as making an important distinction between individuals who have roots in Latin America (including the Caribbean Islands) as opposed to Spanish-speaking European countries. We say diaspora to mean that our participants are individuals who currently live in the United States, including immigrants and US-born individuals.}
 are among the fastest-growing demographic groups in the United States \cite{Zong_2022}, making up ~19\% of the total US population \cite{Zong_2022}, with Pew Research expecting them to represent 14.7\% of voters in the United States by the 2024 US presidential election \cite{Natarajan_2024}. Latinx workers contribute 3.2 trillion dollars a year to the US GDP \cite{Terrill_Harth_2023}, but, on average, have less than a fifth of the wealth of white families \cite{Holzer_William}. On TikTok alone, the “Latino” hashtag has over 70 billion views and over 4 million videos \cite{hashlatino}. However, within online spaces and particularly in the US context, the Latine/Latinx diaspora is marginalized and often demonized. In 2016, after the election of former President Trump with consistent cries to “Build a Wall” and now in 2024 with unprecedented border crossings due to increasing instability in the Latin American world, xenophobic rhetoric against Latinx people is prevalent \cite{Einstein_2023, Gramlich_2023}, particularly on algorithmically curated platforms such as TikTok. Despite this negative content, Latine peoples still flock to these platforms and create their own online spaces within them, with an estimated 1 in 5 adult users on TikTok being Latinx \cite{Valencia_2023}. However, we have not found any HCI research that specifically focuses on Latinx identity formation within the sociotechnical platform of TikTok. 

Without understanding how the large and complex Latine diaspora utilizes and experiences algorithmically enabled media systems like TikTok, this diaspora is left vulnerable to problematic targeted information\cite{newsback}, being the subjects of hateful information, and harassment within these platforms. This project seeks to contribute a necessary bridge between existing identity work in sociotechnical systems and the Latinx population. We build on existing identity space work from queer and Black studies scholars, who have extensively studied how marginalized groups build identities and navigate online spaces through algorithmic marginalization and media representation.

Through a bilingual interview study of 19 individuals and survey of 59, we explore how Latine individuals experience Latine identity, called Latinidad, within sociotechnical systems and through representative media. Our contributions are threefold: (1) an empirical understanding of identity formation as a sociotechnical phenomenon for Latine individuals, (2) an in-depth qualitative study about individual responses to visual media in systems like TikTok, gathered via visual elicitation, and (3) the novel concept of \textit{algorithmic identity compression} as the Latine diaspora on TikTok experiences it.

\section{Related Work}
In this section, we first examine the foundations of Latinidad and existing literature regarding this population in CSCW and HCI. Then, we review existing work on online identity construction and media representation for marginalized populations. Finally, we analyze existing TikTok studies and visual qualitative research methods for our methodology.

\subsection{Latinidad and Latinidad in HCI and CSCW}
Latinidad is an umbrella term for the identity of Latin American-descended persons, as defined in our note on language. There are 33 countries in Latin America, encompassing Central and South America and island nations in the Caribbean \cite{Economic}. Latine people are individuals who have ancestry from Latin America but are commonly classified as “Latino/Hispanic''. This is a misnomer as not every Latin American speaks or has a relationship with Spanish; over 200 million Latine people speak Portuguese \cite{port}. Additionally, there are over 300 languages spoken throughout Latin America, many of them Indigenous \cite{World_Bank_2019}. Regardless of language, Latin American peoples have experienced the repercussions of historic Iberian colonization \cite{Morales_2019}. In the United States, they are racialized and face systemic barriers that Hispanic European immigrants do not \cite{Morales_2019}. Because our study focuses on people of Latin American roots who live in the US, we often say Latine diaspora because this is a diasporic population in the United States. While not all individuals who fall into this category resonate with Latine or Latinx terminology, we have chosen this as a category label for consistency in this paper. The Latine diaspora, given linguistic, racial, and ethnic differences, is uniquely variegated and the myraid of intersecting identities needs to be considered in research. However, we acknowledge that in such a diverse diasporic identity, no label could fully encompass everyone’s identities and experiences.

Current CSCW and CHI literature around Latinx populations focuses primarily on education \cite{forusbyus, confidence, googletech, latinamtraining}, technology design \cite{mobilemoney, womenshealth}, linguistic studies \cite{speechaccent, spanisheducation, editorspanish}, and international research community translation \cite{fosterhci, citationjustice, chicollab, LatinAmCSCW, DoingCSCWLatinAm}. Within education studies, focus has been dedicated to the important task of making sure more Latinx students have access to computing education, given they make up less than 10\% of STEM fields, and that every community has different education needs \cite{Funk_Hugo_Lopez_2022, forusbyus, confidence, googletech, latinamtraining}. Technology design and linguistic studies around and by Latinx researchers focus on technical artifacts that are positioned to unique user needs in Latin America \cite{mobilemoney, womenshealth} and voice enabled technologies that teach or focus on working with Spanish language and accents \cite{speechaccent, spanisheducation, editorspanish}. International research community translation within CSCW and HCI seeks to weigh Latin American contributions equitably as well as understand better research practices for these communities \cite{fosterhci, citationjustice, chicollab, LatinAmCSCW, DoingCSCWLatinAm}. 

Though not centered on HCI, there is previous work focusing on Latinx peoples and TikTok, such as Jaramillo-Dent et al.’s explorative work on the visibility of migrant content on the platform \cite{migranttiktok} and Sued et. al’s study of how app deliverers perform collective narratives on TikTok \cite{appdelivery}. These studies utilize content analysis to understand these specific sub-populations of Latinx peoples and how their posted TikTok content interplays with their offline roles. 

\subsection{Representation of Marginalized Groups in Media}
Recent technological developments have dramatically changed how media is made, distributed, and consumed. Media experts have long described media as systems, notably with Ball-Rokeach and Defleaur’s 1976 media system dependency theory, positing that media must be seen in feedback with societal audiences through systemic frameworks \cite{dependencymedia}. TikTok is a sociotechnical system that enables the creation and sharing of video media through individualized media feeds distributed through an algorithmic recommendation system \cite{socialmediarec, TikTok_2020}, as opposed to other multimedia social networks that focus on networks of followings or text-based content. While these feeds are individualized, similar identity groups often see similar content within this system \cite{infodiet, infodiets2, filterbubbles, echochamber}. 

Each individual’s voluntary choices (i.e., who they follow, what videos they like) and involuntary exposure due to media curations by algorithmic systems influence what they see and therefore contribute to their own understanding of their representation and identity. Media theorists have developed frameworks for understanding how populations have been historically underrepresented in media or stereotyped, reifying real-life social hierarchies 
\cite{MediaSmarts}. Negative mass media representation can result in stereotype advancement and threat for already marginalized groups \cite{Steele_Aronson_1995, stereo2}. Meanwhile, diverse and positive representations of typically marginalized groups have been shown to have positive impacts on and off screen \cite{mediarep, Allyn_2021}. Any widely used media system such as TikTok can dramatically impact both sides of media representation and stereotype threat.

\subsection{Identity Theory Online}
This project draws heavily from the literature of identity theory and formation online, particularly from queer scholarship as well as work regarding racialization in algorithms. We utilize key lenses of identity construction and affirmation in algorithmically controlled online spaces, identity-affirming spaces as well as identity work and flattening.

The experience of identity is highly individual and personal, but it is a social construct structured through membership in and outside social classes and interpreted in systems and cultural norms \cite{Goffman_2022, socialID}. These constructs are also manifested in sociotechnical systems, where racialization and gender categories can be limiting and harmful to those whose identities are not easily contained in preset, often binary, categorizations \cite{gendersee, infodomvig}. 

Online communities have long served as identity-affirming places for individuals. Queer people, people of color, and people going through difficult life circumstances have routinely found affirming and safe  spaces in social media platforms \cite{comingout, safetytrans, classconfessions}. Scholars posit that seeking out identity-affirming spaces in sociotechnical systems is part of doing identity work, where users perform labor to curate routines, community, and content to reaffirm their identities \cite{foryouorforyou, comingout, moderation}. This is a fluid process in which identities shift, and users discover and adapt new ones over time \cite{Ibarra2010IdentityWA}. 

Identities also shift when society and systems do not encompass their complexity. This is seen in Walker and DeVito’s \textit{identity flattening} \cite{moregay} work, where users collapse a multifaceted identity into ‘a less complex presentation of identity that the individuals believe will be more acceptable within a space’ and DeVito’s expansion on this work with transfeminine TikTok creators navigating how TikTok reduces their content categories and identities in the context of content visibility \cite{transfemme}. Users utilize \textit{identity flattening} as a protective measure in response to online harm, to conform to group norms, and to validate identities when invalidated \cite{moregay, classifications}. 

DeVito \cite{transfemme} and Karizat et al. \cite{folktheoryid} define \textit{folk theorization} of identity on platforms, where individuals and groups hypothesize how algorithms function and/or make identity work dangerous and how individuals try to regulate these algorithmic systems \cite{tame}. Finally, Lopez-Leon and Casanova published a case study on theoretical underpinnings of LGBT+ Latinx counterspaces on social media based on the social media usage of their (single) participant \cite{lgbtlatinx}.

However, we found that a Latine perspective on identity experiences in sociotechnical systems, particularly TikTok, was missing. Our work delves into this complex diaspora and focuses on the multiplicity of Latine identities and the specific harms that algorithms can cause by oversimplifying and stereotyping such identities.

\subsection{TikTok in CSCW and HCI}
TikTok has become a massively popular platform in the last 4 years \cite{Saric_2023}. Users may create, upload, or scroll through a feed of videos to consume \cite{tiktokgen}, an algorithmically recommended feed called the “For You Page” (FYP) \cite{TikTok_2020}. 

The full details of the recommendation algorithm of the “For You Page” are the private  property of TikTok. The system learns from users’ behaviors (choosing to watch specific videos, like, comment, etc.) and iterates and evolves the user’s FYP \cite{TikTok_2020}. While following creators makes their content appear on the FYP, users do not control the order or proportions of their feeds, except for the ability to mute certain content with safety features \cite{Pierce_2022}. 

Several empirical and qualitative studies have been conducted on TikTok in HCI venues. There are studies regarding education and informal learning \cite{ghosh2023establishing}, creative collaboration \cite{creativetik}, and political communication on TikTok \cite{partisan}. Most pertinent to our work are analyses of how specific marginalized groups experience TikTok. Much of this scholarship has been spearheaded by queer, Black, and neurodivergent scholars. 

Studies of note that this paper draws from include Simpson and Seeman’s work on LGBT+ users’ routines and experiences with FYPs and \textit{algorithmic exclusio}n \cite{foryouorforyou} and algorithmic curation \cite{tame}, DeVito’s work on trans femme \textit{visibility traps} and \textit{folk theorizations} on TikTok \cite{transfemme}, and Biggs et al.’s work on Queer Farming Utopias on TikTok \cite{farmer}. All of these studies look explicitly at how LGBT+ users navigate TikTok regarding identity affirmation and how they face and combat algorithmic harms from the platform. There are also studies in queer scholarship that center particular harms that this community faces on TikTok from \textit{shadowbanning} \cite{Rauchberg_2022}, a type of algorithmic oppression that suppresses specific videos. \textit{Algospeak}, misspelling or replacing letters in a word, on TikTok has been studied as a mechanism of resilience against shadowbanning \cite{algospeak}. 

Black studies scholars have studied how Black creators are silenced and face harassment on TikTok and other platforms \cite{blackcreators} and how Black creators and users build resilient, joyful, and healing spaces in these platforms \cite{blackharm}. 

We see a variety of identity hashtags promoting storytelling and representation in TikTok studies from neurodivergent scholarship \cite{mentalhealthcomm, laugh}. These studies show the impact and building of identity-affirming spaces via authentic media representation and in-group signaling. Nevertheless, there is still a need for specific studies on the sociotechnical system of TikTok and identity formation for US Latine diaspora populations. Given the underservice to Latine populations in HCI and CSCW research, the lack of Latine representation in STEM roles and therefore decision making regarding the design and development of these systems \cite{Vergara_2023, Arellano_Jaime-Acuña_Graeve_Madsen_2018, PerScholas_2023, Funk_Hugo_Lopez_2022}, and current political events such as the politicalization of Latin American migrants at the US Mexico border and media positioning TikTok as a resource for migrants coming to the US \cite{Turkewitz_2023, Gerber_2023}, we chose to focus on how TikTok’s FYP media curation forms identity spaces for United States diasporic Latine users while simultaneously compressing this identity.

\section{Methods}
This study was conducted under the approval of our university’s Institutional Review Board (IRB). It comprised a survey of 59 participants, followed by 19 semi-structured interviews. Study materials (survey, consent forms, interview proctoring) were available in English and Spanish for all participants. All participants entered this study through a research survey with clear opt-in language and written consent in English or Spanish at the start of the survey. Interview participants signed and returned additional consent forms in English or Spanish before their interviews via email for permission to record and transcribe the interview and for use of their anonymized quotes in this publication. Interviewees also gave verbal consent and had the consent form summarized verbally at the start of the interview before recording started. 

Eligibility requirements for this study were that all participants must be: 18 years or older, live in the United States, understand English or Spanish, use TikTok, see TikTok content related to Latinidad on their FYP, and identify as Latinx. Participants were made aware of these eligibility requirements before starting the survey, and their answers to all requirements were validated before being included in the analysis or invited to participate in an interview. Interviewees had the additional eligibility requirement of needing access to a device that supported Zoom and screen sharing for the interview protocol. 
\subsection{Survey}
The survey served two aims: 1) An entry point for participants to be screened for eligibility and contacted for a research interview and 2) Gathering data about how different Latine individuals used TikTok and content they saw. 
\subsubsection{Participant recruitment and population}\label{3.1.1}
We utilized snowball sampling to recruit participants across a variety of sources. These included: the first author’s social media and personal networks, recruitment emails to listservs of relevant cultural groups at [Anonymous University 1], [Anonymous University 2], [Anonymous University 3], and flyers posted around the general [Anonymous City] area and [Anonymous University 1's] campus. We solicited TikTok users with public accounts identified as Latinx in their biographies via direct messages from their personal TikTok accounts. 

119 participants began the survey, and of those, 65 were completed. 59 of these responses were eligible and used in this study. Ineligible results were determined by location or age ineligibility, spam responses to open questions, and selections of mutually exclusive answers. 

Survey respondents represented 14 Latin American countries, with the majority representing Mexico (n=39), El Salvador (n=7), and Cuba (n=5). 11 participants responded by identifying as Indigenous. Overall, 12 states were represented, with the largest being Washington (n=19) and California (n=12). Most participants were female (n=33), with 23 male participants and 3 nonbinary. 

\subsubsection{Survey design and implementation}
The survey was designed by the first author and reviewed by the second. The survey consisted of 33 questions in total, with a mix of multiple choice (23), Likert rating scale (4), and open-answer (6) questions. The survey asked participants demographic questions specific to their Latine identities and then questions regarding TikTok. Questions covered three themes for participants: their use of the TikTok platform, the content they see on TikTok, and their feelings about this content. The survey was implemented in SurveyMonkey with multiple distribution links to track recruitment results from different recruitment channels. 

\subsubsection{Survey analysis}
Complete and eligible responses were pulled from Survey Monkey and analyzed in private Google Sheets. We followed our IRB’s standards in storing and anonymizing survey data with participants' unique identifiers from Survey Monkey and did not record or store identifying information. We derived descriptive statistics and frequencies of the multiple choice and Likert answers to gain insight into the daily experiences of these users on TikTok. We coded for themes in the open-ended questions which matched the interview protocol to add breadth and depth to the analysis of this population and their experiences on TikTok. 

\subsection{Interviews}
Participants could provide their information at the end of the survey to be recruited for an interview via email from the researchers. Subsequently, if participants responded to this email, following the researchers confirming eligibility of their survey results, they followed the consent procedure described earlier. Interviews and consent materials were offered in English and Spanish. 

\subsubsection{Interview recruitment and sample}\label{3.2.1}
It is worth noting that no sample set can fully represent the entirety of the diverse Latine diaspora. However, we recruited across multiple venues (email, flyers, social media) to try to attain as much diversity as possible in our sample set. Participant demographics represent 7 countries and 9 other Latine identities. In terms of ethnic representation, the majority was Mexican, with 11 of 19 participants identifying as Mexican. We identified 4 distinct lived experiences among our interviewees regarding immigration: People who are immigrants (n=4), US-born individuals who are children of adult immigrants (n=6), people who have experienced or are experiencing being undocumented (n=2), and people who are US-born to US-born parents (n=7). Participants were located in 8 different states, with the majority in Washington (n=5)  and California (n=4). The average age was 25.3, with a minimum of 19 and a maximum of 34. We acknowledge that our participants skewed young; however, this is representative of the age range of TikTok users, where more than half of the platform is aged 18-34 \cite{Winter_2023}. 11 participants identified as female, 8 as male, and all as cisgender. 7 participants identified as part of the LGBT+ community. As Latine is not always a racial category, we asked participants which US census category they utilize and how they racially identified themselves. We also asked what identities, if any, from a list of popular Latine identities informed by literature did participants identify with. This was part of our interview protocol as well. Table \ref{tab: Table 1} has a summary of these identities.

\begin{table}[htbp]
  \caption{List of study participants and relevant identifiers}
  \label{tab: Table 1}
    \begin{tabularx}{\linewidth}{| >{\hsize=.08\hsize}X | >{\hsize=.15\hsize}X | >{\hsize=.08\hsize}X | >{\raggedright\arraybackslash\hsize=.22\hsize}X | >{\raggedright\arraybackslash\hsize=.5\hsize}X |  >{\raggedright\arraybackslash\hsize=.25\hsize}X | >{\raggedright\arraybackslash\hsize=.35\hsize}X | >{\raggedright\arraybackslash\hsize=.35\hsize}X |} 
    \toprule
    ID&Gender&Age&Nationality&Latinx Identities&Location&Census Race Category&Self Racial Identity\\
    \hline

    \midrule
P1&Female&21&Mexico&Immigrant, No Sabo Kid, Dreamer, DACA&Seattle, WA&White&Mestizo, Latino\\
\hline
P2&Female&20&El Salvador, Mexico&Chicana, No Sabo Kid, Child of Adult Immigrants&Seattle, WA&White&Latino\\
\hline
P3&Male&25&Mexico&Afrolatino, Chicano&Los Angeles, CA&Two or more races&Latino\\
\hline
P4&Female&21&Mexico&Indigenous Descent, Chicana&Seattle, WA&White&Latino\\
\hline
P5&Female&22&Mexico&Chicana, Chola, Child of Adult Immigrants&Boston, MA&White&Latino, Hispanic/Latino\\
\hline
P6&Female&25&Mexico&Chicana, Tejana, No Sabo Kid&Philadelphia, PA&Two or more races&Mestiza\\
\hline
P7&Male&24&Brazil&Afrolatino, Chicano&Dallas, TX&Black or African American&Black\\
\hline
P8&Female&29&Peru&Chola, Immigrant&Boca Raton, FL&Two or more races&Latino, Hispanic/Latino\\
\hline
P9&Male&25&Bolivia&Child of Adult Immigrants, Indigenous Descent&Dallas, TX&Black or African American&Black\\
\hline
P10&Female&34&Mexico&Indigenous Descent, Chicana&McAllen, TX&White&Mexican\\
\hline
P11&Male&32&Mexico&Chicano, Cholo&San Francisco, CA&White&Mexican American,  Latino, Hispanic/Latino\\
\hline
P12&Female&22&Mexico&Child of Adult Immigrants&Portland, OR&White&Mestiza, White\\
\hline
P13&Male&25&Cuba&Immigrant&San Jose, CA&White&Hispanic/Latino\\
\hline
P14&Female&19&Ecuador&Child of Adult Immigrants&Ostego, MN&American Indian or Alaska Native &Latino, Hispanic/Latino\\
\hline
P15&Male&32&Mexico&Indigenous Descent, Chicano, Dreamer&Seattle, WA&Two or more races&Latino\\
\hline
P16&Male&25&Bolivia, Cuba&Indigenous descent, Immigrant&San Francisco, CA&White&Hispanic/Latino, Hispanic, White, Latino\\
\hline
P17&Female&26&El Salvador&Afrolatina, Tejana&Houston, TX&Two or more races&Hispanic/Latino, Latino\\
\hline
P18&Male&27&Cuba&Immigrant&San Francisco, CA&Two or more races&Latino\\
\hline
P19&Female&26&Mexico&Indigenous Descent, Chicana, Child of Adult Immigrants&Seattle, WA&American Indian or Alaska Native&Indigenous\\
  \bottomrule
\end{tabularx}
\end{table}

\subsubsection{Interview protocol and moderation}
We conducted 19 total interviews via Zoom, lasting approximately 45 minutes and transcribed by the Zoom software. 17 were conducted in English and 2 in Spanish. Spanish interview transcripts were translated into English by the first author before analysis. Quotes from Spanish interviews are marked as translated in this publication. Participants were compensated with \$30 Amazon gift cards for taking part in the interview process, which lasted an average of 50 minutes. Survey participants did not receive compensation and the average survey time was 5 minutes and 11 seconds.

These were semi-structured interviews with a protocol of descriptive, open-ended questions, followed by a visual elicitation exercise detailed in \ref{interviewanalysis} \cite{Wengraf_2011, glegg}. The questions focused on how the participant’s Latinidad identities had formed and changed throughout their life, then asking them about their TikTok usage, the content they had seen on TikTok, and the experience of their identities on TikTok. For our study, the ability to understand platform usage in the context of lived experience and self conceptualization of Latinidad was particularly important in these interviews, as per other CSCW studies like Simpson and Semaan’s 2020 study on LGBT+ encounters on FYPs that focus on using semi-structured interviews to create life histories about their identities along with platform usage and content \cite{foryouorforyou}.

\subsubsection{Interview analysis}\label{interviewanalysis}
We utilized a constructivist grounded theory \cite{Charmaz_2014} approach to analyze the interviews, particularly around identity formation online allowing for the lived experiences and themes in interviews to construct new theories via relation to existing ones \cite{foryouorforyou, transfemme}. The first author engaged in open-ended coding of the interviews and utilized detailed notes and linked text files to associate open codes with direct quotes from the interview transcripts \cite{Charmaz_2014}. From this, the first author then used thematic analysis \cite{braun} to group these emergent codes into key themes and findings for future relation and analysis to existing theory and the development of our theory of \textit{algorithmic identity compression} (see Section \ref{discussion}). This was an iterative process, with several rounds of coding as the authors met regularly to discuss emerging themes from these iterative analyses. 

To stay consistent with grounded theory approaches, we were careful not to impose categories external to the interview data in these emergent codes and resulting themes. We use the themes to discuss findings in the next section, within the context of the greater survey data and TikTok platform affordances. 

\subsection{Visual Elicitation via Diamond Ranking}\label{diamondrank}
Given the visual nature of TikTok, we chose to incorporate a qualitative visual research methodology via visual elicitation of TikTok content in our interviews to understand how participants conceptualized and related this content to Latinidad. These elicitations served as conversational probes in our interviews to gather further data about how participants thought about this content, in line with qualitative research tradition in using visual media to deepen qualitative insights in interviews \cite{glegg, carlie, graphics}. This conversational data was analyzed with the other interview data with the same methodology explained in \ref{interviewanalysis}.

This was also done acknowledging that we wanted to have conversations around potentially negative media and about the intersection of identities such as race and gender presentation within Latinidad. Visual elicitation has been shown to help with potentially sensitive conversations, as per Dr. Stephanie Glegg’s typology of qualitative visual research methods in interviews in the medical space \cite{glegg}. Additionally, the methodology of showing the participants visual media selected by the research team enabled these conversations without the researcher having to see their FYP, which may be incredibly personal and include items irrelevant to the study. 

The particular methodology utilized in our interviews is the Diamond Ranking Method, a visual research instrument initially seen in the classroom and educational literature to allow children to express feelings regarding different classroom settings or other concepts \cite{clark}. This method has been previously used in HCI literature, particularly at the intersection of HCI and education, in Strohmayer et al's 2015 paper on learning ecologies in homeless populations \cite{hci_diamond}. Diamond Ranking features cells arranged in a diamond shape on one axis from most (the topmost cell) to least (the bottommost cell) of a metric chosen by the researcher \cite{clark}, affording the collection of structured data about how participants conceptualize visual objects or representations with respect to this axis and to one another \cite{clark}. This is illustrated in Figure \ref{fig:1}. An example of this exercise  and media from our interviews is in the Appendix in Figure \ref{fig:3}.

\begin{figure}[h]
  \centering
  \includegraphics[scale=0.35]{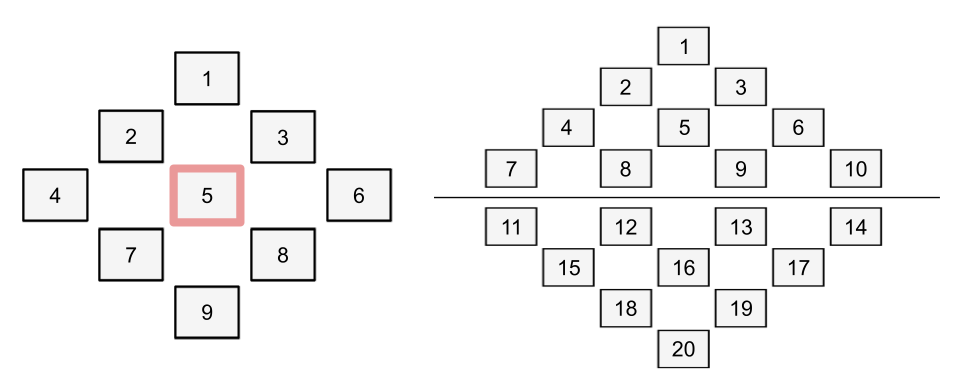}
  \caption{Example of two Diamond Ranking Method instruments, where users would place media objects in the ranked cells. The leftmost diamond contains 9 cells and is not a forced choice, as it has a middle cell highlighted in pink, akin to odd numbered Likert scales. The rightmost is a forced-choice diamond with no middle cell, akin to even numbered Likert scales. Here, the number 1 relates to the most extreme affirmative point of the chosen metric, with each increasing number left to right and top to bottom relating less. }
  \Description{An illustration of two diamond ranking measures. The left is a diamond arrangement of 9 cells with the fifth cell highlighted as the middle cell in pink. The right is the arrangement of 20 cells.}
 \label{fig:1}

\end{figure}

We employ this method in our interviews to understand not only how users reacted to content shown to them in the interview, but also how they organized their knowledge and reactions to content in relation to other content. We adopted Diamond Ranking into a web application for use in the interview, conducting the exercise via Zoom by having the participants open the web application and share their screens. Participants were asked to choose one of their identities (e.g. Mexican) to perform two Diamond Ranking exercises while proctored by the researcher for conversational probing to explain their thought processes. 

The first exercise featured 20 face images and a 12-cell pyramid. This forced participants to leave some images behind and not have a true neutral ranking in their diamond, as per Figure \ref{fig:1}, furthering conversation behind their choices. They were asked to rank these individuals from most to least representative of their conceptualization of the chosen identity (e.g. Mexican). Participants were told this was a thought exercise and were instructed to discuss their logic with the researcher, and reflect on the completed ranking and discuss how their decisions related to their lived experiences or things they’d seen on TikTok. 

Participants then completed a non-forced choice 9-element pyramid of 9 TikTok videos, also proctored by the interviewer. Participants watched all 9 videos and were told to rank them on the same axis of most to least representative of their chosen identity and talk through their ranking logic and reaction to the media with the interviewer. 

The full set of identity options for this exercise was determined from our intake form of our participant demographics, as seen in Table \ref{tab: Table 1}. Each identity had different media collected by the research team and were screened to avoid any inappropriate content, such as violence or sexually explicit content. For the first exercise, the first author searched these identity terms on the TikTok desktop search page and took the first 20 faces from the results on each page via screen cropping. For the second exercise, a random sample of 9 videos 30 seconds or less from this first page was selected.

As this methodology served as a conversational probe and allowed for particular reflections, we did not analyze the rankings themselves, but only the conversation elicited from them.

\subsection{Researcher Positionality}
Both authors identify as women and members of the Latine and Hispanic diasporas in the United States. However, they acknowledge their racialized, linguistic, and citizenship privileges (both authors were born in the United States) within the Latine diaspora, along with the power dynamics of being researchers at a university. While the authors come to this research with cultural and linguistic competencies to analyze this qualitative data, they lack full context for some members of this diverse community, particularly certain members of color, members who have little or no access to the English and Spanish languages, and members who are not US citizens. The first author's positionality also helped with recruitment, given her Latinx networks on LinkedIn and Instagram, but may have been biased towards certain geographic and ethnic identities and other lived experiences in this diaspora. The authors are of different generational demographics: the first author is a Millennial and the second is GenX, giving them different generational reads and perspectives to this data analysis.

\section{Results}
In the following subsections we address our survey and interview findings. We first describe in \ref{4.1} how Latine users experience and use TikTok and in \ref{4.2} users’ folk theorizations about how and why TikTok shows them Latinx content. We then report themes from our visual elicitation exercises with the Diamond Ranking Method; in \ref{4.3} users sort which content creators look and feel Latine to them and in \ref{4.4} we summarize narratives of Latine content from this exercise.

\subsection{Latine users’ content exposure and consumption
}\label{4.1}
\subsubsection{Finding and curating positive, representative spaces}\label{curate} Overwhelmingly, our participants find great joy in their FYPs, particularly regarding the Latine representation and cultural content they see on TikTok. The most frequently encountered content by Latine users in our study was creative content, broader cultural content, and content about Latine lived experiences and identity. We provided in Table  \ref{tab:Table 2} in the Appendix of this paper a full frequency distribution of different categories from our closed codes of our survey and in Table \ref{tab:Table 3} a frequency that combines these closed codes with interview categories of content.

It is worth noting that 10\% of participants do not feel represented at all by their FYPs, with 10\% reporting ‘Almost never’ seeing Latinx people who share their racial identity on their FYPs, particularly Indigenous and Afrolatine participants. 23\% of participants, who identify as LGBT+, do not see Latine people that share their sexuality on their FYP and 7\% of participants do not see their gender identity represented on their FYP. Outside of representation, participants have also encountered negative content, with 42\% of participants encountering false information and 53\% having seen violent content about Latinx people and culture on their FYP.

However, participants still enjoy their overall TikTok FYP experience with 86\% of participants reporting their FYP page makes them feel happy. Interviewees describe feeling represented and included, and they postulate that TikTok is good for representation of the Latine community on and offline. They attribute much of their joy to connecting to shared culture online and seeing others participate in cultural trends. In both surveys and interviews, participants also described feeling joyful learning about their culture and feeling included in the content they see.

\begin{enumerate}
 \item[] “When I see the Samba dance [on a TikTok challenge]...it makes me...feel like I could actually engage in this challenge to express my Latino identity...it gives me the sense of belonging. It brings me joy." - P7
\item []“TikTok has positively impacted my identity by allowing me to learn about my culture's history, new books to read or recipes to try, events happening around the world or near me, etc. This is greatly shaped my viewpoint and cultural experiences of my identity.” - Survey response 
\item[] “I don't think I feel alienated. If anything, I feel included...I really do think it's [TikTok] has helped put us [Latine people] out there more than ever.” - P10
\end{enumerate}

This feeling of connection also came from participants deeming much of their FYP content as authentic, and seeing some of this content with in-group signaling via cultural artifacts and practices, and specific TikTok stylings of these signals, like particular music or sounds over videos. They correlated these types of signals with videos being may by Latine creators for Latine audiences to celebrate and share culture and life moments. During the visual elicitation activity, participants saw videos they deemed to be authentic and discussed how much they valued these types of videos, describing their familiarity with aspects of the videos.
\begin{enumerate}
\item[]“A lot of these were very authentic sounds. I've heard music, I listen to...and like, my Cuban friends who would go back [to Cuba] would have taken those pictures.” - P13.
\end{enumerate}

\subsubsection{Mixed language FYPs of Latine Users}\label{4.1.2}
TikTok’s FYP can be multilingual, showing videos in different languages in the same feed, creating unique experiences for Latinx users. Participants have experienced not just the mixture of English and Spanish content, but also discourse videos in English or Spanish about the the Spanish language and its relationship to Latinidad, with several interviewees (n=9) reporting seeing English content about the Spanish language.

Based on our survey, only 10\% of our participants have a TikTok FYP in one language, with only 8.5\% having only English FYPs. 44\% had mostly English with some Spanish and 36\% report about an even split between Spanish and English. It is important to note that not all our participants have a connection to Spanish, as not all Latin American countries are Hispanic. These participants reported having primarily English content in their FYPs, along with some Spanish. Interviewees also report frequently seeing videos that feature both English and Spanish in the same video. 

91.5\% of all survey participants report seeing content in Spanish on their FYP, grouping them into a Hispanic identity they may not resonate with. In interviews, only 2 participants reported being asked by popup if they knew Spanish, but 18 out of 19 received content in Spanish and English. Spanish language comprehension varies among participants. In our survey, 22\% reported not being fluent or fully comfortable with Spanish, despite being of Hispanic heritage. It is worth noting that this 22\% does not include the 7\% of respondents who have no Hispanic heritage due to being from Brazil, Aruba, or Jamaica.

Participants report mixed feelings about seeing Spanish content and the Spanish language itself. Some participants feel represented when they see Spanish content on their FYP, as they see Spanish as an important community unifier. Other participants did not feel this representation due to feeling like imposters at various points in their life for not knowing Spanish well. They discussed resonating with the “No Sabo Kid” identity that is increasingly popular on TikTok, referring to individuals who do not speak or know much Spanish \cite{Lavin_2023}. However, "No Sabo Kid" and Spanish content are further problematized in Indigenous contexts, with indigenous participants feeling frustrated at the "No Sabo" trend and other interviewees  feeling conflicted about the Spanish language's association with Latinidad, given that it is a colonial language. 

\begin{enumerate}
\item[]“I see Spanish [as]...a sense of pride and unity to the Latino community because it connects us quite a lot." - P9 (Translated from Spanish)
    \item []“Not being able to confidently speak in Spanish has made me question my identity...but Spanish is colonial.” - P2
\item[] “There's a lot of words that...in Spanish, that Indigenous communities say: 'That's kind of a slur to us.'. Everyone says ‘No Sabo’, and Indigenous communities, they'll kind of say: 'I didn't learn Spanish at all, I only know English and my native dialect, please don't call me that.'” - P4
\end{enumerate}

\subsection{Latine users' folk theorization and FYP experiences}\label{4.2}
In a form of folk theorization inquiry, as defined by DeVito \cite{transfemme},  participants hypothesize how an algorithmic system works. Interviewees were asked their hypotheses as to why they are shown Latine content on their FYPs. All (19/19) interviewees report they take deliberate actions to curate an FYP that includes Latine content, but they still see some negative content and conceptualize these as interruptions to their curated pages.

\subsubsection{User actions based on folk theorization}
Participants expressed a strong desire to have Latine content on their FYP and used active measures to try and keep that content. The most popular active measure is following Latinx creators when they see them on their FYPs, something all interviewees report doing and survey respondents also report at a high rate of 83\%. We label active measures users take to influence the algorithm to show them Latine content as \textit{voluntary} user factors. Participants also described \textit{involuntary} factors (e.g. location) they believe TikTok takes into account when showing them this content. We summarize both types of factors in Table \ref{tab: Table 4} in the Appendix.

Participants conceptualize these active measures as a responsibility and method to stay connected to their community and identity. This is largely driven by participants expressing great joy at seeing Latine content from the algorithmic recommendations, as they have not seen it as regularly on other social media. One participant specifically mentioned how the algorithmic recommendations of Indigenous content are helping her reclaim her Indigenous identity and the happiness she felt regarding this. 

\begin{enumerate}
    \item[] “I'm also Indigenous. And I see a lot of [Indigenous videos] on TikTok which is really cool...it means a lot to me when I see something that reflects the Indigenous tribe, or like region where my family is from.” - P4
    \item[]“But when they come up on my For You Page, I do go out of my way...to follow like more Hispanic accounts just to kind of stay in the culture and keep like my language...up to date.” - P13
\end{enumerate}

When probed about what factors they hypothesize TikTok’s algorithmic recommendation system has access to, no participant discussed feeling uneasy about these involuntary user factors being collected by TikTok, such as location tracking to feed them Latine content that reflected their lived experience. In fact, participants often enjoyed being shown very location specific Latine content that they related to and could identify.

\begin{enumerate}
    \item[] “And I think maybe they [TikTok] just track my location. But I've gotten a lot of suggest like “for you” suggestions...to follow people. And I can tell immediately...'Oh, I see those walls, and like you're in Texas’. And that's really fun.” - P6  
\end{enumerate}

\subsubsection{User data hypotheses}
Participants overwhelmingly feel that TikTok has data confirming they are Latinx (6/19), but not always data regarding more granular identities such as Indigeneity or nationalities. Regarding ethnic identities, 13 out of 19 participants felt TikTok had data about their ethnicity and of these, 7 were Mexican. Only 4 participants reported that they suspected TikTok’s algorithm had them associated with a non-ethnic identity, such as “Chicano”, an identity associated with Mexican-Americans \cite{Tatum_2017}. Of these, 2 were Mexican and 2 were not. Overwhelmingly, participants felt TikTok categorized them as Latinx but not anything else, with the exception of Mexican participants who felt strongly TikTok knew they were Mexican.

\begin{enumerate}
    \item[] “I know they [TikTok] know  Latino, but I think they're still trying to guess more.” - P6 
\item[] “I felt like it [the algorithm] already knew I was Mexican, for some reason.” - P12
\end{enumerate}

Some participants report they believe the algorithm is confused or has conflicting data that is changing for them – i.e. showing them one ethnicity for some time, then switching (n = 4). This is largely due to observations that they interact with content that spans identities (i.e. Spanish language content or sports) or interactions with other users of other sub-identities (i.e. Cubans and Mexicans sending TikToks to one another), pushing different content onto their pages. 

\begin{enumerate}
    \item []“I confuse it [the algorithm] so much...it doesn't tell you where I'm from.” - P16
    \item[]“I think they [TikTok] think I am mixed Hispanic, because I get a lot of Mexican content, yes, but I also get like broad general Latinx content. I get some because my partner's Peruvian, so I get a decent amount of Peruvian content. Because one of my closest friends is Guatemalan, I get a lot of Guatemalan content as well.” - P5 
\end{enumerate}

\subsubsection{Negative interruptions and associations to Latine identity spaces}\label{4.2.3}
Our participants believe they have curated positive and often accurate FYPs and affirming identity spaces for their Latinidad. However, several participants (n=7) have experienced interruptions of various degrees of this positive and specific FYP space. 

They theorize this interruption comes from particular content being associated with the content they do enjoy via the recommendation system, but feel very sad and betrayed about this interruption. Three key negative interruption types were identified from interviews: (1) Seeing cultural stereotypes and humor they feel is harmful to Latine communities, (2) news and current events, and (3) seeing Spanish language content when they have not reported knowing Spanish.

Negative, stereotypical humor was a frustrating content interruption experienced by participants on their FYPs, often in the form of skits that belittle Latine femmes, play into other negative stereotypes around Latine people (i.e. drunken, unintelligent Latino men), or other stereotypical content coming from outside of the community. Participants expressed disappointment but not surprise when they encountered this type of content on their FYPs. They theorize that this content is more popular on different parts of TikTok and is recommended to them by related hashtags.

\begin{enumerate}
    \item[] “...every now and then too, my For You Page is like messed up, I will see some where they kind of do like stereotypes. And I don't like...the funny stereotypes when it isn't that actual person...who's from the community....who's like doing these stereotypes of like...the drunk Tio and stuff like that.” - P1
    \item[] “I would categorize that as misogynistic, personally. And I'm sure there's comments on there that are misogynistic as well. Those types of skit videos lead to a lot of negativity.” - P4
\end{enumerate}

Similarly, participants describe news and crises events as further content intrusions recommended to them. Especially when they see negative comments about Latine people, reminding them their feed exists withing a larger platform of other content and features, including harmful viewpoints towards Latine people.
\begin{enumerate}
    \item[] “...there was like some news story I saw where people were saying that like border control officials had shot across the river, and hurt someone on the Mexico side...it's just something I felt really negative about. And going through the comments...there was a lot of like mixed emotions, both positive and negative about the situation. And I was like you would think...going through my For You Page, that you wouldn't see those kinds of comments like that. The fact that you are just kind of like reminded...this really is like an Internet space. It's not just your niche space.” - P5
\end{enumerate}

Participants feel another feature of content recommendation that can be interruptive is when participants who are less comfortable or do not have a relationship to Spanish see Spanish language content. Participants found this invasive and frustrating on TikTok, as they have not signaled to the app that they speak Spanish and may not be fluent in it. This discomfort was not reported for English content, but our two non-English speaking participants had not encountered English content on their FYPs.

\begin{enumerate}
    \item[] "...when I see content that is intensely Spanish...like I speak a little bit of Spanish...I'm not super fluent...So when people just run off with it...that can be a little frustrating." - P6
\end{enumerate}

\subsection{Sorting through who “looks” and “feels” Latine}\label{4.3}
We utilized the Diamond Ranking Method, as explained above in Section \ref{diamondrank}, to understand how interviewees conceptualized Latine TikTok content. Particularly, we wanted to know how representative popular creators were to their concepts of identity. This served as a conversational probe, revealing themes not exposed in other parts of the study. The diamond enabled and facilitated difficult conversations around race, class, featurism, sexism, and stereotypes, and vignettes that would have been difficult to attain without visual elicitation. 

\subsubsection{Organizing subjects as more and less Latine}
When ranking human subjects of popular videos as most to least representative of their identity, participants most heavily relied on facial features and hairstyles (16/19). After phenotypical usage, association with family members (14/19) and friends along with the use of accessories and apparel was most common (14/19). Participants also related \lq emotions' and \lq vibes' they read or felt from the visual media to their lived experiences in this identity group. The full breakdown of the thematic strategies used by interviewees is in Table \ref{tab: Table 5} in the Appendix.

\begin{enumerate}
    \item[] “I think the face structure is very Peruvian....and she has black hair, but she does some highlighting, and that's what we do. We do the highlighting like a lot.” - P8
    \item[] “...[he] kind of reminds me of my brother just when he was younger.” - P2
    \item[] “I associate Cuban people with a lot of expressiveness and...facial reactions behind your emotions.” - P13
\end{enumerate}

All participants acknowledged that any of the faces present could be their identity, acknowledging the wide diversity within the diaspora. However, some faces were left behind more often than others. This was particularly salient for pictures of faces who were identified as ‘Black’ or ‘Afrolatinx’ by participants, who commented that while they knew Afrolatinx individuals were part of the diaspora, they were less represented and discussed this dissonance in the exercise.

\begin{enumerate}
    \item[] “Any of these people can be Mexican...Mexicans can be any color, shape, or size....But I do feel just predominantly what most people consider Mexican, these people definitely fit that more than the other people.” - P10 
    \item[]“I mean, I know there’s like Black Mexicans, I mean Mexico is super diverse, but just from stereotypes...[people] don’t know that.” - P15
\end{enumerate}

Afrolatinx participants (n=3) in our study ranked these images higher, but also acknowledged that these images are not what most people envision when they think of Latine peoples and they purposefully have had to seek out Afrolatinx content on TikTok as they did not find it was recommended to them as often. Participants discussed how they have faced anti-Black racism and bias within the Latine diaspora and have not always seen their identities represented in Latine spaces, and have had their Latine identities questioned due to their Blackness.
\begin{enumerate}
    \item[] “You know how we [Latine people] have a whole lot of issues that arise when it has to do with skin color. So most times people might have a different perspective and belief about you.” - P9 (translated from Spanish)
    \item[] "People are always like ‘Are you both your parents Salvadorian? Are you sure?’ And I'm like: ‘I'm sure.’” - P17
\end{enumerate}

\subsubsection{Default Latino == Mexican}
A key theme that surfaced across the Diamond Ranking of videos and pictures was many participants spoke about Mexicans being the “default Latino” in the United States and how they see this being repeated in TikTok videos as well. Many participants discussed this in relationship to appearances of individuals in the exercise. They felt overall Latinx people and content are associated with being Mexican, and Mexcians are overrepresented on and offline. Many non-Mexican participants have been assumed to be Mexican by non-Latine people and in turn felt unseen and not reprensented in and out of Latinx discourse.

\begin{enumerate}
    \item[] "[When] we think about Latinidad..It's very Mexican centric." - P2
    \item[] “I think what most people think when they think of Latinx people, they think about like Mexicans...when I was small they [schoolmates] used to just think that I was Mexican...which it did kind of irritate me.” - P14
\end{enumerate}

However, many non-Mexicans have found community with Mexican individuals over commonalities, such as language and food and seeing that content on TikTok. Participants discussed that this community and content feel more familiar to them than much white American content, and add to their feeling on belonging on their FYP and in real life.
\begin{enumerate}
    \item[] “...Mexican community...through the language, and food...lots of things bridge the [cultural] gap, or make me feel like I'm part of the larger sort of group.”  - P13 
\end{enumerate}
Seven Mexican participants discussed how TikTok knew about their ethnicity and suggested that TikTok may be primed to guess Mexican over other ethnicities, acknowledging their privilege in being frequently represented in Latine content and spaces, including TikTok, where they frequently see content that is specific to Mexicans or Mexican Americans. 
\begin{enumerate}
    \item[]"I feel like I have the privilege to feel seen by a lot of Latinx community in the US. I don't think it's all-inclusive of the greater Hispanic community." - P5
\end{enumerate}
However, some Mexican participants (n = 6) discussed feeling targeted by this default and visibility by non-Latine people, such as assumptions around Mexicans being undocumented and uneducated, while other communities may not be as predominantly associated with these negative stereotypes due to less visibility. 
\begin{enumerate}
    \item[]“...like a lot of people associate [being undocumented] with Mexican, specifically ...that's hard to navigate.” - P1
\end{enumerate}

\subsection{Evaluating popular and limiting narratives in Latine content}\label{4.4}
Along with discussing which creators were seen as more or less Latine, participants also ranked content videos. Common themes and tensions were identified in this portion of the interview, with two key narratives emerging: (1) Negative stereotypes of Latina women and their intelligence, (2) associations with Latino men and manual labor.

Many recipients felt \lq stuck' in these narratives and similar ones, which they felt portrayed Latine people as always being poor with limited options. While they acknowledged that many Latine people have less wealth than white households, they still crave more diverse representation about Latine lived experiences. Some participants even felt that seeing these limiting narratives repeatedly in social media could feed into predestination, with people feeling they have less options and and not pursuing alternatives due to this representation.

\begin{enumerate}
    \item[]“But there's literally always going to be more to us than those two socioeconomic labels.” - P4 
    \item[]“There's a narrative of hard work which implies physical labor. And I think as Mexicans and Latinos, we're really proud of that. But I think that also gets us stuck in not believing that we can do other things....Not necessarily bigger things. That might be exactly what you want to be doing. But knowing that you're capable of doing whatever it is that you want to do and that you see other races...or other groups of people do.” - P15 
\end{enumerate}

\subsubsection{Sexy, dumb Latinas}\label{hotcheeto}
In many of the Diamond Ranking Method datasets, there were skit videos portraying femme presenting Latine individuals or individuals portraying them in negative ways. Particular content of note for participants was the “Hot Cheeto Girl” stereotype enacted in one of these skit videos, a negative archetype of Latina women that displays them as rude, stupid, and disruptive, reifying existing biases against women of color \cite{Froio_2023}. 

Participants all responded negatively to the “Hot Cheeto Girl” skit, and other content related to it. They reflected on having seen this content on their own FYPs and other social media networks and felt many of these videos portray Latinas as dumb and overly sexual. Participants discussed how this stereotype had effected them and people they had known negatively in real life, but also highlighted how so many Latinas in their lives counteracted this stereotype while having some of the traits that are over-exaggerated in these videos. Participants underscored wanting this narrative to end and be able to claim some of these traits of femininity and humor for their own without negative assumptions about their intelligence for liking spicy food and having acrylic nails.

\begin{enumerate}
    \item[] “They're never portrayed as being intelligent. It's just like screaming and...eating Hot Cheetos...it just sucks. I feel like I was never taken seriously in school because of like stereotypes like this”. - P12

    \item[] “I think it definitely could be harmful because it just like downplays, a lot of a person's like inner attributes...especially downplays like intelligence. And a lot of my friends I know that have these Hot Cheeto-like personalities, are actually some of the smartest people that I know. They're all in grad school right now...these stereotypes about those personalities downplays that.”  - P5

\end{enumerate}

\subsubsection{Manual laborer Latino men}
Participants felt torn regarding the association between Latino men and manual labor (construction, landscape, etc). For many participants (n = 7) this type of elicitation reminded them of loved ones, particularly fathers. Participants expressed a desire for this type of labor to be valued and respected, and that they felt pride in the hard work that this labor requires being embedded in their cultures and lives. However, like the "Hot Cheeto Girl" stereotype, they do not want other traits like intelligence to be extrapolated from this association, and more representation for the vast array of Latine lived experiences beyond manual labor, without discounting it.

\begin{enumerate}
    \item[] “I feel like sometimes people think that Mexicans being considered hard workers or just construction or just laborers is derogatory. And I'm never gonna think that that's derogatory. I embrace it. And I love the fact that people think we're hard workers and...I know a lot of people don't like seeing, you know, brown people in hard hats...but...we're not all construction workers....Every year the population with degrees that are Hispanic grows and grows. I love that. But also I take pride in the trade and the blue-collar side of of our people. And you know the people picking our food, deserve all the respect in the world. So I will never see that as a negative.” - P10 
\item[] “...a lot of us in the community tend to work...hard labor jobs...my dad also has a vest like this one.” - P1

\end{enumerate}

\section{Discussion}\label{discussion}
Through our empirical interview and survey findings, thematic analysis, and existing literature, we define 4 key tensions the Latine diaspora experience in sociotechnical systems, and how they  manifest on TikTok: 1) Machismo and gender based discrimination, 2) Linguistic assimilation, 3) Intersectionality of immigration and representation in sociotechnical systems, and 4) Racialization in sociotechnical systems. We evaluate these tensions in the context of social and cultural theory and affordances of sociotechnical systems, particularly TikTok. From these tensions and affordances, we build and present our theoretical contribution of \textit{algorithmic identity compression} in  Section \ref{aicdiss}, and show how TikTok and similar sociotechncial systems result in algorithmic identity compression for Latine diaspora members. 

\subsection{Tensions in content and sociotechnical systems for Latine diaspora}
\subsubsection{Machismo and gender-based discrimination}\label{mach}
\textit{Machismo} is a unique cultural manifestation of misogyny in the Latine diaspora. Many marginalized groups experience misogyny differently given interlocking identities and forces. An example of this is \textit{misogynoir}, coined by Moya Bailey, to describe the unique intersection of misogyny and racism that Black women face \cite{Bailey_2021}, and has been built upon by Turner et al. in understanding how it manifests in sociotechnical systems \cite{playbook}. Machismo has been attributed to Spanish colonialism and studied at great length by Chicanx scholars in how it impacts family dynamics and the life experiences of Latinos (male Latine people) \cite{machismo}. Recent literature has shown that younger Latine people are moving toward less stringent patriarchal roles and expectations, as this rigidity is linked to negative social and physical health outcomes for Latinos \cite{machismorecent}. Our interviewees echoed this sentiment, expressing how they have seen machismo content on and offline that harms Latinos, limiting their attitudes about their potential career and educational paths, and in expressing their emotions. Participants wanted to see alternative representations and for their Latine cultures to stop claiming and reinforcing machismo.

\begin{enumerate}
    \item[]“...men felt like they couldn't be emotional, because, you know, they have to provide, they can't focus on anything else...we shouldn't claim...machismo, as part of Chicano culture...we should change it.” - P19
    \item[]  “But then it also can be negative, because I have seen like videos where it will be maybe one of the construction worker...and they'll say stuff like, ‘No, I'm Mexican. I don't go to university. I don't go to college.’” - P11
    \end{enumerate}

Machismo enables a culture that is unsafe for Latinas and Latinx femmes \cite{chicana}. This is both by devaluing Latinas as inferior to men and forcing the female gender role into \textit{Marianismo} \cite{APA}, a gender role of submissiveness and chastity. In the United States, Latinas must navigate a double standard of this submissive role with over sexualization thrust upon them from media tropes such as “Spicy Latinas” \cite{Lopez_2021}. The “Spicy Latina”, an archetype that stresses the exotic sexuality of Latinas, often with a ‘rough’ background and confrontational personality, and has been linked to stereotype threats of Latina women in modern media and in real life \cite{percept, eva}. This stereotype has evolved into sexist online memes like the “Hot Cheeto Girl”,  \cite{Froio_2023}, creating content that devalues Latinas as unintelligent, inferior, and disruptive. These "Hot Cheeto Girl" skits appeared in our study, as discussed in \ref{hotcheeto}, with participants finding them sexist and problematic.

These skit videos are not only harmful in content, but also in scale, as they which have great potential to go viral. A brief search of hashtags of “\#HotCheetoGirl” and “\#LatinasBeLike” shows these hashtags are massive and linked to several trending audios \cite{hashlatina}. TikTok presents a formulaic platform where this content can quickly be reposted, stitched, and recreated as it gets popular, reifying existing stereotypes at massive scales.

This devaluation and stereotype threat have real-life consequences. In the United States, Latina and Black women are more likely to experience violent crimes than white women, with Latinas making up 37\% of missing women and 42\% of homicides in 2021 and 2022 \cite{lacivil}. Latinas also have less economic mobility, making only 55 cents on every dollar that a white man makes \cite{California}, often being stereotyped into difficult and low-earning labor associated with their gender roles. 

Participants felt that other video content was very sexualized towards Latinas. While there was no sexually explicit content in our study, there was content with sexual undertones. A key media element was a video about “Spicy Latinas” wearing sexy pajamas to bed for their white American boyfriends, which many interviewees connected to the underlying sexism of the "Hot Cheeto Girl" and explicitly felt frustration and discomfort regarding this video.

\begin{enumerate}
    \item[] “That's a Latina stereotype...I think that was very pervy.” - P8
\end{enumerate}

These trends feel like transgressions, as many participants associate feminine beauty and sexuality with positive cultural aspects, like dance and dress. Brazilian Carnaval is an epitome of this, with many scholars and activists citing it as an example of reclaiming sexuality and gender expression for both femme and masc presenting Latine peoples \cite{carnival, brazil}. 

\begin{enumerate}
    \item[] “She looks very beautiful. She is sexy...it's a cultural stuff. And anyway, and you watch this person you're definitely going to, you know, feel more, close to culture because of the dance...” - P7
    \item[] “She looks amazing, beautiful! I can say she represents all the girls... because I see some dance there I can see some of the beauty there. I can see the Vibe” - P3 (translated from Spanish)
\end{enumerate}

Participants felt that social media and systems reinforce the negative stereotypes and impose hypersexuality on Latinas without consent. On TikTok, participants felt this manifested in popular videos they had seen on the app. Indeed, across TikTok and other social media platforms, results for \#Latina are full of sexually explicit videos that are near the very top of the search page \cite{Noble_2018}. This type of harm against women of color has been studied at length by scholars such as Safiya Noble investigating search algorithms proliferating sexualized and explicit images of Black women \cite{Noble_2018}. A study by Ghosh and Caliskan confirmed this bias and over-sexualization of women of color, particularly citing Latina examples, has proliferated from online spaces to the machine learning models behind text-to-image generators \cite{ghosh-caliskan-2023-person}. 

\subsubsection{Assimilation}
Participants discussed a feeling of assimilation and loss of culture, both from the colonial loss of Indigenous languages from Iberian colonization but also from having to assimilate to English upon moving to the United States. The loss of Indigenous languages in Latin America is stark; the World Bank estimates 26\% of indigenous languages are at risk of disappearing \cite{World_Bank_2019}. This is a result of a long history of colonialism that consistently marginalized, discriminated against, and harmed Indigenous communities throughout the Americas \cite{ghost}. This has resulted in Indigenous families having to learn colonial languages and even hide their Indigenous heritage due to discrimination \cite{ghost}. By the time these families moved to the United States, their Indigenous knowledge was gone and many participants in our study were unsure of the details of their Indigeneity and had lost the language entirely. 

\begin{enumerate}
    \item []“I know my dad's side comes from Indigenous background. But they don't really know, where or how.” - P15
    \item[]“My grandma, she speaks it [Indigenous language] fluently, and I think Spanish is actually her second language. My dad...can speak a bit. But...they [other people] would treat you lower...If you had any resemblance to like anything Indigenous...back when I was growing up there, it was pretty bad.” - P16
\end{enumerate}

Some participants report reclaiming this Indigenous identity and having important learning experiences on TikTok as described in \ref{curate} where participants feel that algorithmically curated Indigenous content brings them joy and that TikTok is a safe and affirming space for them to consume this content. Some participants have also used hashtags to learn about Indigenous identities, but while TikTok supports Spanish language codecs, there is no sense of Indigenous language in TikTok or video codecs for many other platforms \cite{TikTokforDevelopers}, showing further language loss even as the World Bank and other institutes make efforts to preserve these languages \cite{World_Bank_2019}.

Along with Indigenous language loss, participants often discussed Spanish language loss from their family’s assimilation in United States. Participants spoke of how their parents did not teach them Spanish, prioritizing English, often to avoid discrimination. Some families saw a loss of their indigenous language and Spanish within only two generations. This assimilative experience left some participants discontent with TikTok’s association of Latinidad with Spanish language videos, but finding community in the  “No Sabo Kid” trend discussed in \ref{4.1.2}.

\subsubsection{Immigration tensions and privileges}
Another tension within the Latine diaspora that manifests within TikTok and other sociotechncial systems is that of immigration status. In \ref{3.2.1} we define the 4 types of immigration experiences our participants had, with 8/19 being immigrants themselves, including 2 who had experienced being undocumented. The rest of our study were US-born children of adult immigrants (6/19), as are 50\% of Latinx US-born people \cite{Hispanic_ResearchCenter_2022} or US-born Latine children to US-born parents (7/19), having families that have been in the US for generations, with regional identities like Chicano and Tejano that serve as  foundations of Latinx culture in those areas \cite{Tatum_2017}. 

Both groups experience fluidity and tensions around their immigrant or citizenship identities, but in different ways and systems. For US-born Latine participants, much of this tension came from experiences in college education. Meanwhile, immigrant participants discussed the labor, precarity, and isolation of navigating the immigration system.

44\% of Latinx students are the first in their family to attend university \cite{Flores_2021}, where many realized they were part of a minority in a way they were not in their home community. They also learned that within the Latine community, there were wider socioeconomic realities, echoing back to their frustration with media and platform representation painting Latine people as only lower income as seen in \ref{4.4}. This countered their experience of meeting adult Latine immigrants who were more economically privileged and did not grow up within the disproportionate racial and economic discrimination experienced by the US Latinx diaspora \cite{Holzer_William}. This manifests within higher education with the underrepresentation of US-born Latinx professors, who make up less than 10\% of Latinx engineering professorships in the United States \cite{Arellano_Jaime-Acuña_Graeve_Madsen_2018}, while adult Latinx immigrants comprise the other 90\%, adding a particular tension within these groups in many other systems, such as who goes on to achieve influence in various STEM sectors. This was a source of frustration for many participants, who described these tense encounters during their university experiences.
\begin{enumerate}
    \item[]"...[people would] keep saying: 'Oh, you wouldn't understand. You're not like an immigrant.' I was like: 'Well, also, you don't understand being like a poor brown person in America.'" - P6
\end{enumerate}

However, many US-born Latine people, including our participants, understand the privilege that comes with US citizenship and the precarity of the immigration experience, particularly in being undocumented. An estimated 35\% of the Latine diaspora in the US are undocumented \cite{Millet_Pavilon_2022}, and 1 in 4 US-born Latine children have a parent who is undocumented \cite{Hispanic_ResearchCenter_2022}. Over 500,000 Millennials and Generation Z are DACA (Deferred Action for Childhood Arrivals) recipients, individuals who were undocumented as children but can attain specific work, educational, and citizenship possibilities \cite{migrationpolicy.org_2023}. Despite this, there is a large stigma around DACA and immigration status in the US and Latine diaspora. Many immigrant participants discussed the fluidity of this, the challenges of navigating the US immigration system throughout their lives. A particular experience was when achieving more privilege and security, through measures like greencards, many expressed grief and guilt in losing immigrant experience communities and camaraderie.

\begin{enumerate}
    \item[]“But once I got my green card, I felt like I wasn't able to go anymore. And I didn't tell them but I just stopped going [to an undocumented support group].” - P1
\end{enumerate}

\subsubsection{Racial tensions and privileges} 
There is a vast range of intersectional identities within the Latine diaspora. Participants honed in on this during their Diamond Ranking exercises, reflecting on how different individuals within the diaspora were treated both in and outside of the group, particularly based on race and anti-Black sentiments. 

Whiteness has unique underpinnings in the Latine diaspora, deriving from colonial histories like the \textit{castas} system that placed white colonizers at higher social strata than Indigenous and African descended peoples. This resulted in practices such as \textit{blanqueamiento} by which families strategically tried to marry into whiter ones to advance their positionality \cite{bodypolitic, ghost}. While the \textit{casta} system is no longer de jure, colorism and anti-Black discrimination remains in Latin America and the US \cite{charles}. This results in unique discrimination against  members of the Latine diaspora, especially those of African descent or with darker skin \cite{Noe-Bustamante_Gonzalez-Barrera_Edwards_Mora_Hugo_Lopez_2021}. This colorism is reaffirmed in media, which affects other systems such as image-to-text generators, which favor lighter skin tones and oversexualizes or demeans Latine people of color \cite{ghosh-caliskan-2023-person}. 

While participants regularly noted they were aware that anyone could be Latine, both their lived experiences and Diamond Ranking exercises echoed existing biases towards Afrolatinx and Indigenous people, as discussed in \ref{4.3}. These biases contribute to an exclusive model of what a Latine person looks like – associating Latinidad with lighter-skinned people. This reifies and adds to representation power within media consumption, as most users consume a never-ending stream of this content and engage with it. This can have repercussions within the creator economy. If Afrolatine users are not considered representative by many Latine users, algorithms may suppress this content if people they classify as Latine engage with it less. This cycle continues and reifies who is not shown as Latine online.

Many Latine people identify as white racially due to mixed ancestry or being white-passing. All participants who who identified as white or white-passing (7/19) in our study acknowledged their whiteness, or proximity to whiteness, grants them privilege as they navigate through society, but they did not want to distance themselves from their Latine identities.  Additionally, whiteness was not seen as the most representative of Latinidad by participants in the ranking exercises. But while many participants do not conceptualize themselves as white nor Latinidad as white, they often default to marking “White” in various forms and databases because of Latine not being a racial category and because of the loss of their Indigeneity, not wanting to fill out forms as Indigenous.

\begin{enumerate}
\item[]"When I still fill out the forms, I just put like, white just because that's what I've been told to do." - P1
\item[]"I feel a lot of people within my culture either put like Native American or white, depending on how much they identify with their Indigenous identity." - P5
\end{enumerate}

Latine people are no one race, having gone through multiple waves of colonization by European settlers and interactions with the transatlantic slave trade \cite{ghost, bodypolitic, Morales_2019}. Latine people being encoded as "white" in sociotechnical systems is an emergent property that ignores this complex history and can reify exclusionary practices. This can create unintended consequences in systems, ranging from the type media recommended on platforms like TikTok, but also in ignoring the racialized experience of many Latine people in important databases used to make real-world decisions.

\subsection{Algorithmic identity compression}\label{aicdiss}
These experiences of our participants affirm the Latine diaspora is not monolithic and that affordances of systems like TikTok do not encompass the multifacted realities of this group. Subsequently, the Latine diaspora is simplified to unrepresentative categories and associations that can cause harm through reification of biases, problematic content, and erasure of intersecting identities in data systems. These harms must be reckoned with by the designers of these systems

To account for this phenomena, how it arises, and its ramifications in sociotechnical systems, we propose the concept of \textit{algorithmic identity compression}. We define this as the simplification, flattening, and conflation of intersectional identities to a larger grouping due to the affordances of an algorithmic system and existing societal biases at the cost of losing critical cultural community features that are deemed unnecessary by these systems and designers of them. We argue that Latine users experience algorithmic identity compression within TikTok and other sociotechnical systems, such as aforementioned biasing of AI systems against Latinas or defaulting to whiteness in racial data categories.

This theory builds upon concepts such as Simpson’s \textit{algorithmic exclusion} on TikTok for LGBT+ users \cite{foryouorforyou}, Donna Haraway’s concept of \textit{informatics of domination} \cite{infodomvig}, Walker and DeVito’s \textit{identity flattening} concept, and critical scholars' work on identity classification limitations and instabilities in database and information systems \cite{gendersee, gendersocialmedia, comingout, misgenderingmachines}. These aforementioned works explore how sociotechnical systems reify Western and hegemonic binary categories and do not make space for non-normative or marginalized identities, and enact harm onto these individuals, even in their own spaces. We add to this work with algorithmic identity compression in that this compression is imposed by interlocking choices in systems and how features are compressed into monoliths with the loss of critical cultural data.

We purposefully use the metaphor of \textit{compression} informed by how \textit{image compression} shrinks files within an ‘acceptable’ threshold  \cite{Gonzalez_Woods_2019}. We argue that sociotechnical systems are setting this threshold very low for marginalized experiences. Furthermore, \textit{image compression} also reduces file size to make the content faster to deliver \cite{Gonzalez_Woods_2019}. In a sociotechnical system like TikTok where more content being recommended and engaged with provides more revenue, this speed and shrinkage are ingrained as an acceptable tradeoff for important metadata, such as not moderating explicit imagery associated with \#Latina. But why do more complex identities need to be shrunk for faster mass content delivery? And what essential societal information is lost with this compression? And given the paucity of Latines in technical positions \cite{Vergara_2023, Arellano_Jaime-Acuña_Graeve_Madsen_2018, PerScholas_2023}, who gets to set this threshold of algorithmic identity compression through the system?

TikTok enacts algorithmic identity compression of Latinidad through 3 key affordances: (1) Linguistic complexities via singleton video codecs, (2) ethnic and gender identities through hashtag conflations, and (3) dominant negative stereotypical narratives through trending videos and audio. These tensions  are summarized in Table \ref{tab:Table 6} in the Appendix.

In most digital platforms, including TikTok, videos are labeled, or encoded, in only one language \cite{TikTokforDevelopers}, with some languages, like Indigenous ones, excluded entirely. This singularity leaves little to no room for the multilingual videos that are so common in the Latine diaspora, such as English videos that feature Spanish phrases, erasing this multilingualism and associated labor. Any linguistic nuance is compressed to one language in the video codec, most likely based on a classification of the audio which may not be able to identify less served languages such as Indigenous languages, leading to them not being codified, algorithmically suppressed, and not reaching their audiences, obscuring content around Indigeneity. By extension videos codified in one language are recommended by algorithms to individuals associated with that language – regardless of their personal comfort or association. 

\begin{figure}[h]
  \centering
  \includegraphics[width=0.6\linewidth]{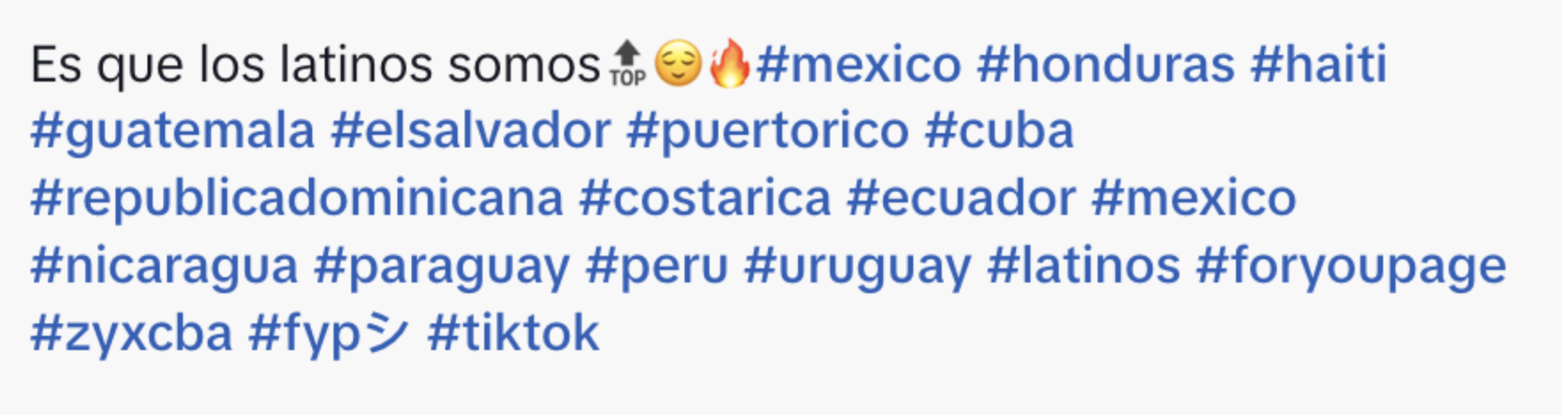}
  \caption{Example of hashtag accumulation under common \#latino videos featuring a variety of ethnicities for more engagement.}
  \Description{The bottom of a TikTok video showing \#latino conflated with other hashtags of Latin American countries.}
    \label{fig:2}
\end{figure}

Hashtags are another mechanic of algorithmic identity compression for Latinidad on TikTok, particularly for ethnic identities. As hashtags become more popular, more videos are grouped under them. Additionally, on a platform like TikTok, we see many, many hashtags being used on videos, such as “\#Latino” followed by \#Mexico, \#Cuba, \#PuertoRico, \#Bolivia, etc. – making sure to capture many sub-identity hashtags with the main \#Latino \cite{hashlatino}. An example of this is visible in Figure \ref{fig:2}. TikTok’s hashtag structure obscures individual Latine identities into the greater umbrella of “Latino” or “Latinx” through how hashtags and the FYP algorithm are structured. This reinforces biases around Latinidad being a monolithic experience and aims content at as large of an audience as possible, with no embedded knowledge or acknowledgment of the group. This can directly cause negative interruptions due to malicious actors realizing they can hijack these large community hashtags for negative or violent imagery.

We illustrated the effects of compression of “Latina” identities with sexualized images in \ref{mach}, which is an example of lossy compression, likened to \textit{lossy image compression} where images cannot be restored. The association of hypersexuality and Latinas has become so ubiquitous across many sociotechnical systems and representations that while it is enabled by platforms via mechanics like hashtags, it is a manifestation of offline biases, including creators and audiences.

Influencers are pressured to chase trends and reach as many views as possible, often by recreating these popular negative narratives. These narratives, normalized via societal biases and so easily shown via platform affordances, are often accepted and engaged with. Thus, Latine influencers continue to uphold these hegemonic narratives about and within their own communities to build audiences, often including non-community member audiences. In this structure, non-Latine influencers are also incentivized to appropriate and uphold these trends.

No sociotechncal system will ever capture the nuances of every lived experience within a diverse group, but we desire systems to be critical in who, and by what metrics, makes compression decisions. There are some pragmatic improvements for platforms, such as increasing moderation of explicit sexual content of Latinas or introducing more language codecs and one to many database relationships to foster multilingualism and Indigenous language promotion. We hope this scholarship around algorithmic identity compression will be a useful tool to design such improvements and in examining this phenomena for other pluralist groups in other sociotechnical systems.

\section{Limitations and future work}
Like any qualitative empirical study, no sample can be or is meant to be generalizable across a diverse group like the Latine diaspora. Our study sample was limited in age range, ethnicities, political viewpoints, and racial diversity. 

The age range of our sample, while in line with over half of TikTok users being 18-34 \cite{Winter_2023}, excluded perspectives of older Latinx persons. This impacted findings, such as reflecting on more recent educational experiences and on how the social concept of race has changed, not representing the mixed-race tension of the second author’s GenX experience.

Interviewees represent 7 countries and 9 other Latine identities, with a majority Mexican representation (11/19) and only 3 interviewees identifying as Afrolatinx. We suspect political viewpoints and snowball sampling contributed to selection bias, as evinced by so many of the interviewees' awareness of white privilege and intersectionality. While we offered Spanish interviews, most participants chose English, which may have biased findings such as multilingualism as in \ref{4.2.3}. Additionally, offering the study in more languages, like Portuguese, could have brought more perspectives. 

Also, as we specifically recruited people who interact with TikTok regularly and live in the US, another selection bias was introduced, as participants with positive experiences with TikTok would be more likely to self-select for this study, perhaps leading to the overall positive view interviewees had of their FYP feeds.

Future work could target recruitment to include more diverse perspectives, such as more nationalities, racial identities, political viewpoints, and age ranges. Future studies focused on Latine individuals that faced harm on TikTok may reveal insights and address some of the selection bias this study could have faced. Additionally, these methodologies and the application of \textit{algorithmic identity compression} to other identity groups could strengthen this theoretical framework.

\section{Conclusion}
This study explored for the first time how members of the US Latine diaspora experience identity formation on TikTok through sociotechnical and media representation lenses. Our findings show that the Latine experience is unique and worthy of further study, and while users curate joyful feeds on TikTok, its algorithmic affordances create tensions through content and affordances that reify existing biases. 

Participants desire to see quality, nuanced, and diverse representation of Latine people online and via TikTok. This brings them joy. It is in online spaces that they are currently seeing this representation, but they are being forced to accept it at a cost of transgressions such as explicit content associated with their identities and violent news stories that intrude upon their curated feeds. Participants expressed a common desire to dismantle the existing negative stereotypes this type of recommended media reifies and bring nuance and authenticity to their lived experiences online. We describe this reification and lack of nuance as \textit{algorithmic identity compression}. 

We found algorithmic identity compression of the Latine diaspora is specifically enacted by TikTok through 3 key affordances: (1) Singleton video codecs, (2) hashtag conflations, and (3) trending videos and audio (see \ref{aicdiss} and Table \ref{tab:Table 6}).

As algorithmic systems advance in power and reach, it is imperative that the nuances of identities of marginalized populations are captured and that the negative stereotypes around them are moderated. This work reveals how interconnected these negative stereotypes are to not just media representation, but other informatic data categorizations and sociotechnical systems and how they have acute and real-world impacts on Latinx individuals. 

\section{Acknowledgments}
We are sincerely grateful to our participants, without whom this work would not exist. We are grateful to our department, for supporting this work via the [Anonymous University] [Anonymous Department]'s doctoral research grant.
\bibliographystyle{ACM-Reference-Format}
\bibliography{2_bib}

\appendix

\section{Tables and Figures}
\begin{table*}
  \caption{Frequencies of closed codes of categories on surveys}
  \label{tab:Table 2}
    \begin{tabular}{p{0.7\linewidth} | p{0.1\linewidth} | p{0.1\linewidth}l} 
    \toprule
   Content Category&Participants&Percentage\\
   \hline

   \midrule
    Content about Latinx culture - music, visual art, dancing, book, movie, or other media reviews, food, sports, traditions, and other cultural displays&52/59&88.1\%\\
    \hline

    Creative content around Latinx - skits, POVs, and other freeform creative content surrounding one of the above categories or with Latinx coded characters in fantasy worlds&43/59&72.9\%\\
    \hline

Content about Latinx identity and lived experience - adoption from Latin America, Latinx families, Latinx cultural exchanges, Latinx identity, Latinx stereotypes, etc&40/59&67.8\%\\
\hline

Content about Spanish language, but not specific to Spain - Spanish language instruction, content about bilingualism and Spanish, etc&36/59&61.0\%\\
\hline

Content about Latinx locations - Latinx neighborhoods in the US, Latin American countries, Latin American points of interest (i.e. monuments, memorials, etc)&33/59&55.9\%\\
\hline

Content about Latinx historical or current events - politics, natural disasters, historical analyses, current events (i.e. migrations), laws, etc&32/59&54.2\%\\

  \bottomrule
\end{tabular}
\end{table*}
\begin{table}[]
\end{table}
\begin{table}[]
\end{table}
\begin{table}[]
\end{table}

\begin{table}
  \caption{Open and closed codes of categories}
  \label{tab:Table 3}
    \begin{tabular}{p{0.3\linewidth} | p{0.5\linewidth} | p{0.1\linewidth}l} 
    \toprule
   Open emergent code&Related closed code from survey&Interviewees\\
   \hline

   \midrule
   Sports&Content about Latinx culture&2/19\\
   \hline

   First generation college content&Content about Latinx identity and lived experience&3/19\\
   \hline

   Indigenous content&Content about Latinx historical or current events, Content about Latinx identity and lived experience&4/19\\
   \hline

   News/political events&Content about Latinx historical or current events&6/19\\
   \hline

   Fashion&Content about Latinx culture&7/19\\
   \hline

   Food&Content about Latinx culture&9/19\\
   \hline

   Spanish language content&Content about Spanish language&9/19\\
   \hline

   Informational Latinidad content&Content about Latinx identity and lived experience&10/19\\
   \hline

   Acting out/POVs/Skits&Creative content around Latinx &11/19\\
   \hline

   Comedy/Memes&Creative content around Latinx&11/19\\
   \hline

   Day in the life/story times&Content about Latinx identity and lived experience&13/19\\
   \hline

   Music and Dance&Content about Latinx culture&14/19\\
   
  \bottomrule
\end{tabular}
\end{table}

\begin{flushleft}
    \begin{table}
  \caption{Grouping of voluntary and involuntary user factors that interviewees hypothesize lead to Latinx content being on FYPs}
  \label{tab: Table 4}
    \begin{tabular}{p{0.5\linewidth} | p{0.3\linewidth} | p{0.1\linewidth}l} 
    \toprule
    User factor&Voluntary or Involuntary factor&Interviewees\\
    \hline

    \midrule
    Uses reposting, stitching, and other creator tools&Voluntary&4/19\\
    \hline

    Posts own content about Latinx experiences&Voluntary&7/19\\
    \hline

    Uses hashtags to look up content&Voluntary&8/19\\
    \hline

    Uses search bar to look up Latinx content&Voluntary&9/19\\
    \hline

    Intentionally follows Latinx creators’&Voluntary&19/19\\
    \hline

    Intentionally likes Latinx content&Voluntary&19/19\\
    \hline

    Location of device and self&Involuntary&7/19\\
    \hline

    TikTok eavesdropping on them&Involuntary&9/19\\
    \hline

    Having a Hispanic name&Involuntary&4/19\\
    \hline

    Friends sending them Latinx content on the TikTok app&Involuntary&11/19\\
  \bottomrule
\end{tabular}
\end{table}
\end{flushleft}

\begin{table}[]
\end{table}
\begin{table}[]
\end{table}
\begin{table}[]
\end{table}

\begin{table}
  \caption{Breakdown of thematic strategies for image organization by participants }
  \label{tab: Table 5}
    \begin{tabular}{p{0.4\linewidth} | p{0.15\linewidth}l} 
    \toprule
    Method of organization&Interviewees\\
    \hline

    \midrule
    Visual facial features/hair&16/19\\
    \hline

    Association of family members/friends&14/19\\
    \hline

    Used apparel/accessories&14/19\\
    \hline

    Used perceived role of people&12/19\\
    \hline

    Stereotypes/media&12/19\\
    \hline

    Used context of background of images&9/19\\
    \hline

    Known influencer/figures&5/19\\
  \bottomrule
\end{tabular}
\end{table}
\begin{table}[]
\end{table}
\begin{table}[]
\end{table}
\begin{table}[]
\end{table}

\begin{table}
  \caption{Compression affordances and examples on TikTok}
  \label{tab:Table 6}
    \begin{tabular}{p{0.25\linewidth} | p{0.3\linewidth} | p{0.35\linewidth}l} 
    \toprule
    Compression Effect&TikTok Affordance&Example\\
    \hline

    \midrule
    Linguistic identity compression&Singleton video codecs&A video in Spanish and English only being codified in English and being shown to any Latine assigned user\\
    \hline

    Ethnic and gender identity hashtag conflations&Having large amounts of hashtags possible on videos&A video featuring \#latino and then hashtags of over 10 Latin American countries, but the video only pertains to Mexican Americans\\
    \hline

    Dominant, negative steoretypical narratives&Trending videos and hashtags that are pushed to more FYPs&The \#HotCheetoGirl is a very popular hashtag that keeps getting recommended and reposted in this media ecosystem.\\

  \bottomrule
\end{tabular}
\end{table}

\begin{figure}[h]
  \centering
  \includegraphics[scale=0.35]{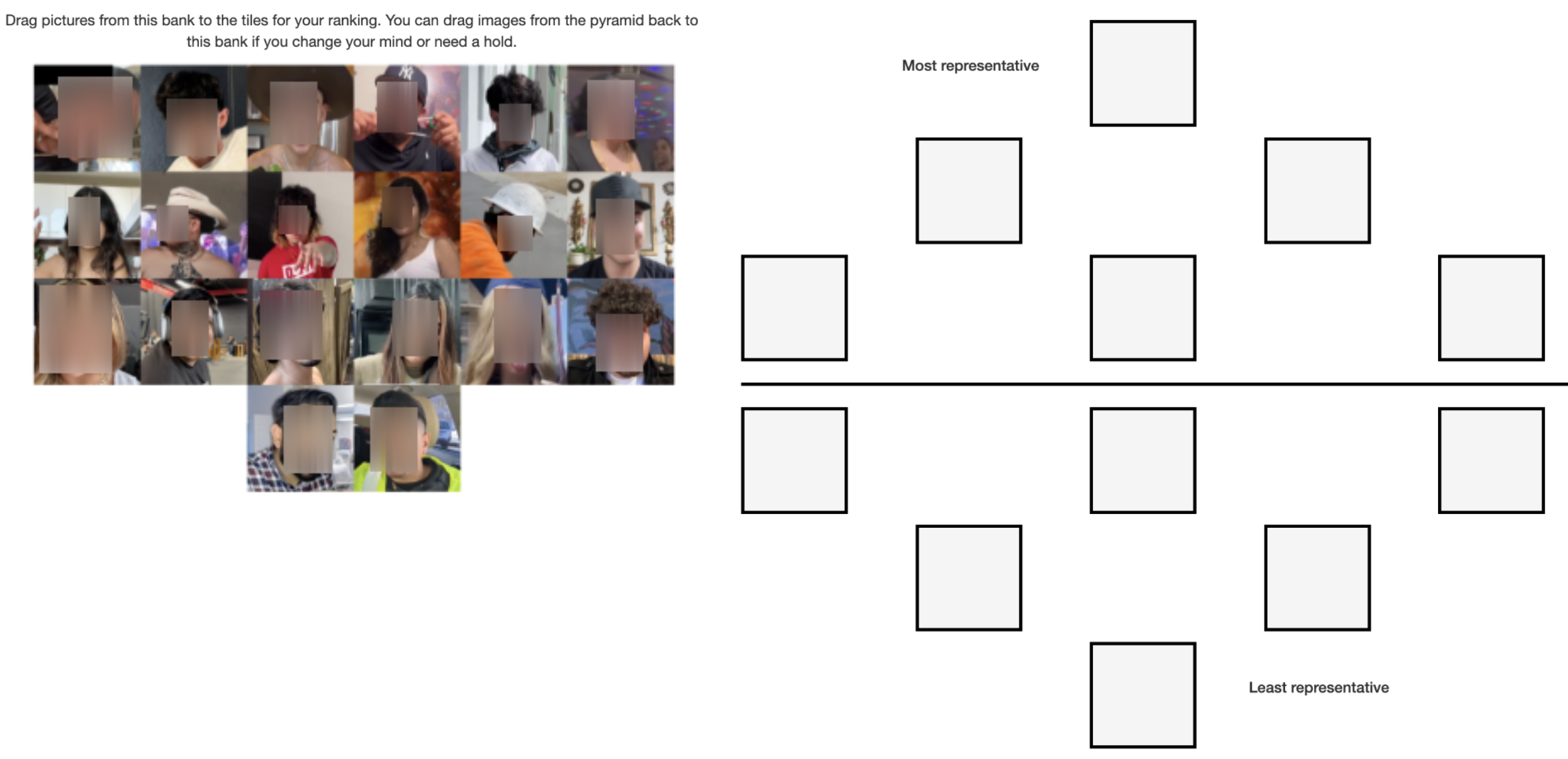}
  \caption{Example of a Diamond ranking exercise used in the interviews described in \ref{diamondrank}. In this activity, participants were shown photos of creators (blurred here for privacy) and were asked to drag them into boxes based on most to least representative based on their conceptualization of their Latine identity. This activity served as a conversation probe during the interview.}
    \label{fig:3}

\end{figure}

\end{document}